\documentclass [12pt]{article}
\usepackage {graphicx}
\usepackage{bm}
\usepackage{amsmath}
\usepackage{amssymb}
\usepackage{cite}
\usepackage{amsmath, amssymb}
\usepackage[amsmath, thmmarks]{ntheorem}
\theoremstyle{plain} \theoremheaderfont{\normalfont\bfseries}
\theorembodyfont{\itshape}
\newtheorem{theorem}{Theorem}[section]

\theoremstyle{plain} \theoremheaderfont{\normalfont\bfseries}
\theorembodyfont{\itshape}

\theoremstyle{plain} \theoremheaderfont{\normalfont\bfseries}
\theorembodyfont{\itshape}

\theoremstyle{plain} \theoremheaderfont{\normalfont\bfseries}
\theorembodyfont{\itshape}

\theoremstyle{plain} \theoremheaderfont{\normalfont\bfseries}
\theorembodyfont{\itshape}

\theoremstyle{plain} \theoremheaderfont{\normalfont\bfseries}
\theorembodyfont{\itshape}

\theoremstyle{plain} \theoremheaderfont{\normalfont\bfseries}
\theorembodyfont{\normalfont}
\newtheorem{example}[theorem]{Example}

\theoremstyle{plain} \theoremheaderfont{\normalfont\bfseries}
\theorembodyfont{\normalfont}

\theoremstyle{plain} \theoremheaderfont{\normalfont\bfseries}
\theorembodyfont{\normalfont}

\theoremstyle{plain} \theoremheaderfont{\normalfont\bfseries}
\theorembodyfont{\normalfont}

\theoremstyle{plain} \theoremheaderfont{\normalfont\bfseries}
\theorembodyfont{\itshape}

\begin{document}

\title{Holder-extendible European option: corrections and extensions}

\author{Pavel V.~Shevchenko\\
\\\footnotesize{The Commonwealth Scientific and Industrial Research Organisation}\\
\footnotesize{Locked Bag 17, North Ryde, NSW, 1670, Australia}\\
\footnotesize{e-mail: Pavel.Shevchenko@csiro.au}\\
}

\vspace{2cm}

\date{\footnotesize{1st version: 28 September 2010, this version 19 September 2014}}
\maketitle

\begin{abstract}
\noindent Financial contracts with options that allow the holder to
extend the contract maturity by paying an additional fixed amount
found many applications in finance. Closed-form solutions for the
price of these options have appeared in the literature for the case
when the contract underlying asset follows a geometric Brownian
motion with the constant interest rate, volatility, and non-negative
``dividend" yield. In this paper, the option price is derived for
the case of the underlying asset that follows a geometric Brownian
motion with the time-dependent drift and volatility which is
important to use the solutions in real life applications. The
formulas are derived for the drift that may include non-negative or
negative ``dividend" yield. The latter case results in a new
solution type that has not been studied in the literature. Several
typographical errors in the formula for the holder-extendible put,
typically repeated in textbooks and software, are corrected.

\vspace{0.5cm} \noindent \textbf{Keywords:} exotic options,
extendible maturities, holder-extendible option, geometric Brownian
motion.
\end{abstract}
\pagebreak

\section{Model}
Financial contracts with options that allow the holder to extend the
contract maturity by paying an additional fixed amount found many
applications in finance. The European option with extendible
maturity (written on the underlying asset $X_t$) can be exercised by
the holder on a decision time $T_1 $ using strike $K_1 $. The holder
may also exercise the option later at some maturity $T_2>T_1$ using
strike $K_2 $ by paying an extra premium $A>0$ at time $T_1$. Denote
the value of this option at time $t\le T_1$ as
$Q(X_{t},t;K_1,K_2,T_1,T_2)$ and we want to find the fair value of
this option at zero time $t=T_0=0$. At time $T_1 $, the payoffs for
the holder-extendible call and put are

\begin{equation}\label{holder_call_payoff_eq}Q_C (X_{T_1 },T_1;K_1,K_2,T_1,T_2) = \max \left(X_{T_1 } - K_1 ,C(X_{T_1 }
,T_1;K_2,T_2) - A,0\right)
\end{equation}
\noindent and
\begin{equation}\label{holder_put_payoff_eq} Q_P (X_{T_1 } ,T_1;K_1,K_2,T_1,T_2 ) = \max \left(K -
X_{T_1 } ,P(X_{T_1 } ,T_1;K_2,T_2) - A,0\right)
\end{equation}

\noindent respectively. Here, $C(X_t,t;K,T)$ and $P(X_t,t;K,T)$ are
the standard European call and put at time $t$ respectively for spot
value $X_t$, strike $K$ and maturity at time $T$; that is, their
payoffs at maturity are $\max(X_T-K,0)$ and $\max(K-X_T,0)$
correspondingly.

Applications of these options include extendible options on foreign
exchange, non-dividend and continuous dividend yield stocks, real
estate, bonds, etc. For example, the standard holder extendible
option in foreign exchange (FX) allows the holder to extend the
maturity of FX vanilla option by paying an extra premium; option on
real estate often allows the option holder to extend the contract
expiry date by paying additional amount to the option writer. In
general, any contract that may involve rescheduling payments could
be viewed as including an option with extendible maturity.
Closed-form solution for these options were presented in Longstaff
(1990), Haug (1998, p.48), Chung and Johnson (2011), and Chateau and
Wu (2007) in the case when the underlying asset $X_t$ follows a
geometric Brownian motion with the constant drift and volatility.
For more details and applications on the extendible options, the
reader is referred to the above-mentioned publications.

In this paper, we consider geometric Brownian motion model with the
time-dependent drift and volatility which is important for practical
applications. Specifically, we assume that the underlying asset $X_t
$ follows the risk-neutral stochastic process

\begin{equation}
\label{riskneutralprocess_eq} dX_t = X_t \mu(t)dt + X_t \sigma
(t)dW_t ,
\end{equation}

\noindent where $W_t$ is a standard Wiener process, $\sigma (t)$ is
the instantaneous volatility, $\mu(t)=r(t) - q(t)$ is the
risk-neutral drift, $r(t)$ is the risk-free domestic interest rate
and $q(t)$ is some known continuous function of time (hereafter
referred to as ``dividend"). For example, this model is often used
for pricing a holder-extendible option on a foreign exchange rate,
where $q(t)$ corresponds to the foreign interest rate; in the case
of option on dividend paying stock, $q(t)$ corresponds to the
continuous dividend yield. Longstaff (1990) and Chung and Johnson
(2011) consider the case of zero ``dividend" $q(t)=0$; Haug (1998)
and Chateau and Wu (2007) consider the case of non-negative dividend
$q(t)\ge 0$; also constant drift and volatility are assumed in these
studies. In this paper, we allow for negative $q(t)$ (e.g. negative
foreign interest rate in the case of FX options) leading to a new
solution type that has not been considered in the literature; also
the drift and volatility are allowed to be time dependent.

Under the process (\ref{riskneutralprocess_eq}), the joint
distribution of $\ln X_{T_1}$ and $\ln X_{T_2}$, given $X_0 $, is a
bivariate normal distribution with
\begin{eqnarray}
\label{mean_cov_eq}
\begin{array}{l}
\mbox{E}[\ln X_{T_i } \vert \ln X_0 ] = \ln X_0 + \int_0^{T_i }
\left(r(\tau ) - q (\tau ) - \frac{1}{2}\sigma ^2(\tau )
\right)d\tau ,\quad i = 1,2; \\ \\
\mbox{Cov}[\ln X_{T_i } ,\ln X_{T_j }
\vert \ln X_0 ] =
\int_0^{\min(T_i,T_j)} {\sigma ^2(\tau )d\tau },\quad i,j=1,2.\\
\end{array}
\end{eqnarray}

Then, according to the standard option pricing methodology, a fair
price of the holder-extendible option at $t=0$ is a conditional
expectation

\begin{equation}
\label{optionprice_general_eq} Q(X_0 ,0;K_1,K_2,T_1,T_2) = e^{- r_1
T_1} \mbox{E}[Q(X_{T_1 } ,T_1;K_1,K_2,T_1,T_2 )\vert X_0 ],
\end{equation}

\noindent where $Q(X_{T_1 } ,T_1;K_1,K_2,T_1,T_2 )$ is given by
(\ref{holder_call_payoff_eq}) and (\ref{holder_put_payoff_eq}) for
the holder-extendible call and put respectively. The above
expectation can be calculated using (\ref{mean_cov_eq}) and integral
identities (see Appendix) in closed form as demonstrated in the
following sections. We derive option price formulas both for the
holder-extendible call and the holder-extendible put; the formulas
are presented for the cases of non-negative and negative
``dividend". Note that, some of the conditions and formulas
presented in the previous literature, e.g. Longstaff (1990) and Haug
(1998), have erroneous errors subsequently repeated in textbooks,
other papers and software; these are fixed in this paper.

%Unfortunately there are several typographical errors in the
%published formula Longstaff (1990, equation 12) for the
%holder-extendible put. These are subsequently repeated in textbooks,
%other papers and software. For example, formula for the
%holder-extendible put in Haug (1998, equation 2.15, p.48) still has
%a typographical error. Chung and Johnson (2011) presents formulas
%for extendible options in the case of an arbitrary number of
%extensions. The general formula for the holder-extendible put in
%Chung and Johnson (2011) will reduce to the correct expression in
%the case of one extension date. However, it is not an easy exercise
%to extract this simple case from the general formula. Chung and
%Johnson (2011) points to the typographical errors in Longstaff
%(1990, equation 12) but fails to list all corrections required.

\section{Notation and definitions}
Hereafter the following notation and identities are used.
\begin{itemize}

\item Model parameters
\begin{eqnarray}
\begin{array}{l}
q_{ij} = \frac{1}{T_j-T_i }\int_{T_i }^{T_j } {q(\tau )d\tau
},\quad r_{ij} = \frac{1}{T_j - T_i}\int_{T_i }^{T_j } {r(\tau )d\tau },\quad \mu_{ij}=r_{ij}-q_{ij}. \\
\\
\sigma_{ij}^2 = \frac{1}{T_j-T_i }\int_{T_i}^{T_j } {\sigma ^2(\tau
)d\tau },\quad\mbox{and}\quad\rho = \frac{\sigma_{01} \sqrt {T_1 }
}{\sigma _{02} \sqrt {T_2 }
}\\
\end{array}
\end{eqnarray}
for $T_i<T_j$ and $i,j=0,1,2$.
\item Transformation functions
\begin{equation}
\label{transform_func}
\begin{array}{l}
g_1(y)=\frac{\ln (y / X_0 ) - \mu _{01} T_1 + \frac{1}{2}\sigma
_{01}^2 T_1 }{\sigma _{01} \sqrt {T_1 } },\quad
\widetilde{g}_1(y)=g_1(y)-\sigma _{01} \sqrt {T_1 },\\
\\
g_2(y)=\frac{\ln (y / X_0 ) - \mu _{02} T_2 + \frac{1}{2}\sigma
_{02}^2 T_2 }{\sigma _{02} \sqrt {T_2 } },\quad
\widetilde{g}_2(y)=g_2(y)-\sigma _{02} \sqrt {T_2 }
\end{array}
\end{equation}
and their inverse
\begin{equation}
\label{invtransform_func}
\begin{array}{l}
g^{-1}_1(y)=X_0\exp(\mu _{01}T_1 - \frac{1}{2}\sigma _{01}^2
T_1+\sigma
_{01}\sqrt{T_1}y),\\
\\
g^{-1}_2(y)=X_0\exp(\mu _{02} T_2 - \frac{1}{2}\sigma _{02}^2
T_2+\sigma _{02}\sqrt{T_2}y).
\end{array}
\end{equation}

\item Critical values of the underlying asset defining exercise
regions on a decision time are denoted as $\{I_1,I_2,I_3\}$ for the
holder-extendible call and $\{J_0,J_1,J_2\}$ for the
holder-extendible put.

\item $N(\cdot)$ and $N_2(\cdot ,\cdot ;\rho )$ are the
standard normal distribution and the standard bivariate normal
distribution with correlation $\rho$ correspondingly; their
densities are denoted as $n(x)$ and $n_2(x,y;\rho )$ respectively.
$M_2(a,b,c,d;\rho)$ is the probability of the standard bivariate
normal density with correlation $\rho$ for the region
$[a,b]\times[c,d]$, and $M(a,b)$ is the probability of the standard
normal density in the interval $[a,b]$ that can be expressed through
$N(\cdot)$ and $N_2(\cdot ,\cdot ;\rho )$ as given in Appendix.

 \item The standard European call and put prices at time $T_i$ with maturity $T_j>T_i$:
 \begin{eqnarray}
&&C(x ,T_i; K ,T_j)=x
e^{(\mu_{ij}-r_{ij})(T_j-T_i)}N(d_1)-Ke^{-r_{ij}(T_j-T_i)}N(d_2);\\
&&P(x ,T_i; K ,T_j)=Ke^{-r_{ij}(T_j-T_i)}N(-d_2)-x
e^{(\mu_{ij}-r_{ij})(T_j-T_i)}N(-d_1);\\
&&d_1=\frac{\ln(x/K)+(\mu_{ij}+\frac{1}{2}\sigma^2_{ij})(T_j-T_i)}{\sigma_{ij}\sqrt{T_j-T_i}};\quad
d_2=d_1-\sigma_{ij} \sqrt{T_j-T_i}.\nonumber
\end{eqnarray}

\item To compare the calculus with Longstaff (1990), one has to set
\begin{equation}
\label{comparison_setting_eq}\sigma_{01}=\sigma_{02}=\sigma_{12}=\sigma,\quad
\mu_{01}=\mu_{02}=\mu_{12}=r_{01}=r_{02}=r_{12}=r.\end{equation}
\end{itemize}

\noindent The choice of some notations is dictated by the purpose of
easier comparison with already published results rather than
simplicity and generality.

\section{Holder-extendible call}
\label{para:solution} 
\begin{center}
\begin{figure}[p]
\centerline{\includegraphics[scale=0.65]{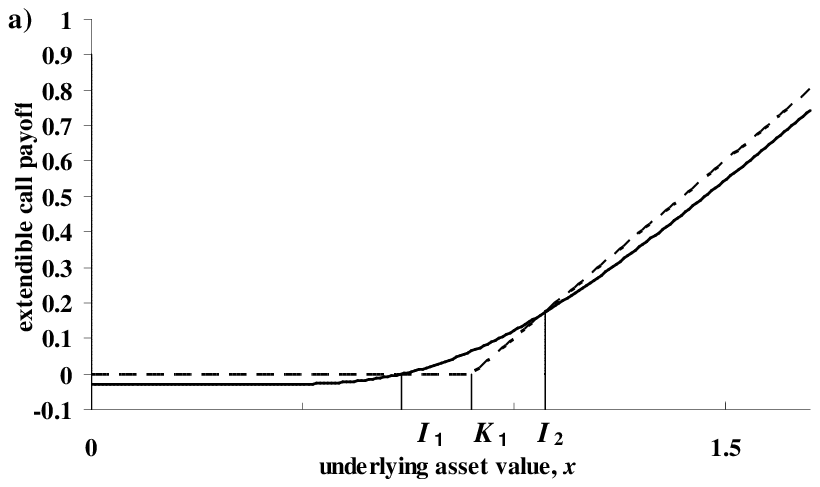}\hspace{0.2cm}\includegraphics[scale=0.65]{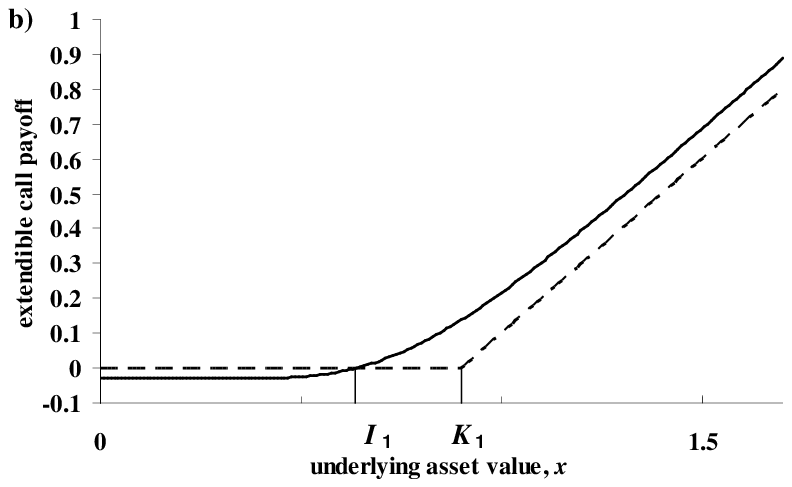}}
\vspace{0.5cm}
\centerline{\includegraphics[scale=0.65]{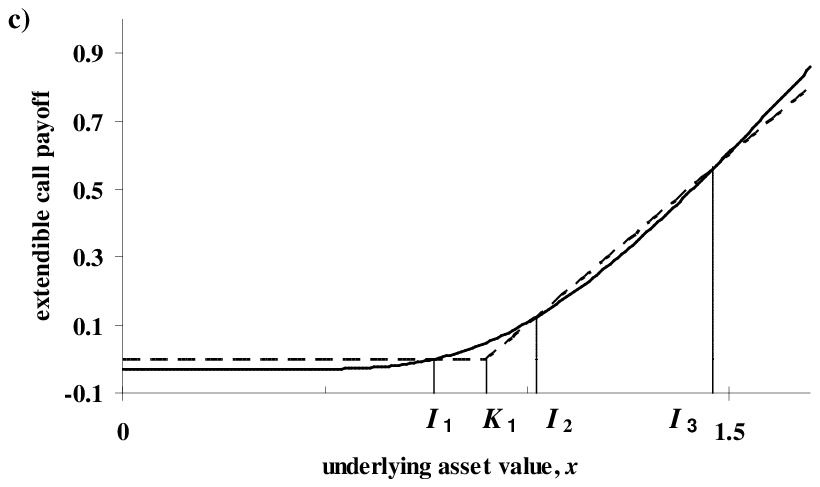}\hspace{0.2cm}\includegraphics[scale=0.65]{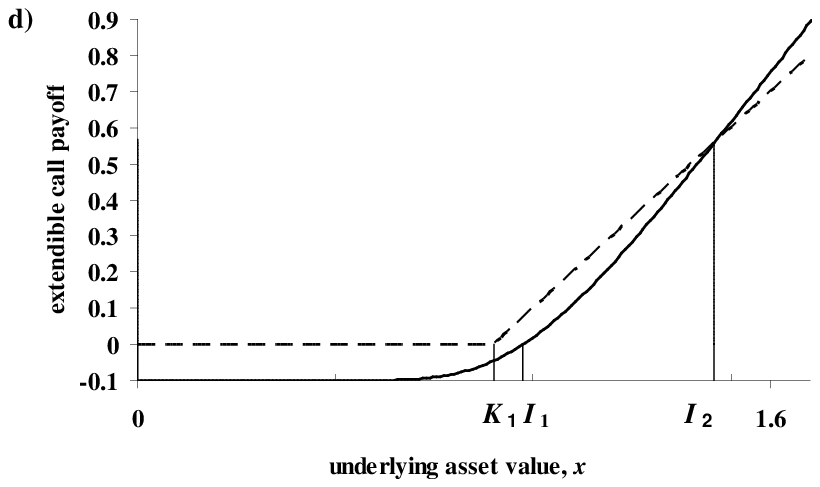}}
\vspace{0.5cm}
\centerline{\includegraphics[scale=0.65]{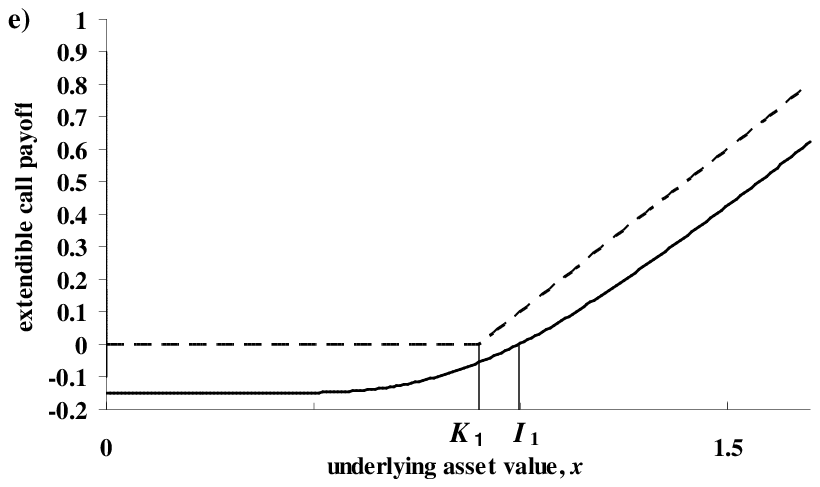}}
\caption{Possible holder-extendible call payoffs on a decision time
$T_1$. The payoff is determined by choosing the largest value
between solid line $C(x,T_1;K_2,T_2) - A$ and dashed line $\max
\left(x - K_1 ,0\right)$.} \label{Call_fig}
\end{figure}
\end{center}

The decision at $T_1 $ to extend or exercise
the call option is determined by comparing two risky payoffs
\begin{equation}
C(X_{T_1 } ,T_1;K_2,T_2) - A\quad\mbox{and}\quad \max \left(X_{T_1 }
- K_1 ,0\right);
\end{equation}
\noindent and choosing the largest payoff. If the first payoff is
larger then the option is extended; otherwise it is exercised when
$X_{T_1 } > K_1$ or expires worthless when $X_{T_1 } \le K_1$; for
an illustrative example, see Figure \ref{Call_fig}. Note that the
standard European call $C(x ,T_1; K_2 ,T_2)$ is calculated at time
$T_1$ for maturity $T_2$.

Denote the region of $X_{T_1}$ values where the option is extended
as
\begin{equation}\Omega_C=\{x\ge 0: C(x ,T_1;K_2,T_2) - A > \max \left(x - K_1
,0\right)\} \end{equation}

\noindent and the region where it is exercised as

\begin{equation}\overline{\Omega}_C=\{x> K_1: x - K_1
\ge C(x ,T_1;K_2,T_2) - A \}.\end{equation}

\noindent For all other values of $X_{T_1}$, the option expiries
worthless. Then, using transformation of $X_{T_1}$ and $X_{T_2}$ to
the random variables $Z_1=g_1(X_{T_1})$ and $Z_2=g_1(X_{T_2})$ from
the standard normal distribution, the today's price
(\ref{optionprice_general_eq}) of the holder-extendible call can be
written as

\begin{eqnarray}
\label{call_intergal_solution_general_eq}
 &&Q_C (X_0 ,0;K_1,K_2,T_1,T_2) = e^{ - r_{01} T_1 }\int_{ - \infty }^\infty  \max
\left(C(x_1,T_1;K_2 ,T_2) - A,x_1 - K_1
,0\right){n({z}_1)d{z}_1} \nonumber\\
&&\hspace{2cm}=e^{ - r_{02} T_2 }\int_{{x}_1\in\Omega_C}
 {d{z}_1\int_{g_2(K_2)}^\infty {(x_2  - K_2 )n_2({z}_1,{z}_2;\rho )d{z}_2} } \nonumber\\
 &&\hspace{2.5cm} - e^{ - r_{01} T_1 }A\int_{{x}_1\in\Omega_C}n({z}_1)d{z}_1
+  e^{ - r_{01} T_1 }\int_{ x_1\in\overline\Omega_C} n({z}_1)(x_1 -
K_1 )d{z}_1.
 \end{eqnarray}

\noindent Here $x_1=g^{-1}_1({z}_1)$ and $x_2=g^{-1}_2({z}_2)$ are
functions of ${z}_1$ and ${z}_2$ as given by
(\ref{invtransform_func}).

 The regions
$\Omega_C$ and $\overline\Omega_C$ can be determined using solutions
(critical asset values) of nonlinear equations
\begin{equation}
\label{call_crit_equation1_eq} f_1^C(x)=C(x ,T_1; K_2 ,T_2) - A=0,\quad x\ge 0\\
\end{equation}
\noindent and
\begin{equation}
\label{call_crit_equation2_eq} f_2^C(x)= C(x ,T_1;K_2 ,T_2) - x +
K_1 - A=0, \quad x > K_1.
\end{equation}

\noindent These can be solved numerically using e.g. the
Newton-Raphson algorithm combined with the standard bisection
algorithm to avoid numerical difficulties when corresponding
derivatives are close to zero.

The first equation $f_1^C(x)=0$ has one solution, denoted as
$x=I_1$. It is bounded as

$$Ae^{q_{12}(T_2-T_1)}\le I_1 \le
Ae^{q_{12}(T_2-T_1)}+K_2e^{-\mu_{12} (T_2-T_1)}. $$

\noindent The second equation $f_2^C(x)=0$ may have two, one or no
solutions depending on the option characteristics (strikes,
maturities, model parameters) that will determine the today's option
price. If exist, the solutions will be denoted as $I_2, I_3$; Figure
\ref{Call_fig} illustrates possible cases. Below we consider two
distinct cases of non-negative and negative ``dividend", i.e. the
cases $q_{12}\ge 0$ and $q_{12}<0$ respectively. This is because if
$q_{12}\ge 0$ then $f_2^C(x)=0$ may have one or no solution; and if
$q_{12}<0$ then $f_2^C(x)=0$ may have two solutions. Note that
Longstaff (1990) formulas correspond to zero ``dividend''
$q_{12}=0$.

All conditions, listed in Sections \ref{Call_nonnegdiv_sec} and
\ref{Call_negdiv_sec}, on option characteristics to determine
solution type can easily be proved using the facts that the European
call price $C(x,T_1;K_2,T_2)$ is a continuous and increasing
function of $x$, and its first derivative
\begin{equation}\label{CallDelta_eq}
\Delta_C(x)=\frac{\partial
C(x,T_1;K_2,T_2)}{\partial
x}=e^{-q_{12}(T_2-T_1)}N\left(\frac{\ln(x/K)+(\mu_{12}+\frac{1}{2}\sigma^2_{12})(T_2-T_1)}{\sigma_{12}\sqrt{T_2-T_1}}\right)
\end{equation} is positive. It is important to note that  $0\le\Delta_C(x)\le 1$
when $q_{12}\ge 0$; however, if $q_{12}< 0$ then $\Delta_C(x)>1$ is
possible.

\subsection{Non-negative ``dividend''}\label{Call_nonnegdiv_sec}
Consider the case of non-negative ``dividend", $q_{12}\ge 0$.
\begin{itemize}
\item If $I_1\ge K_1$, then the call is never extended (i.e. $f_2^C(x)=0$
has no solutions for $x>K_1$) and thus
$$Q_C(X_0 ,0;K_1,K_2,T_1,T_2)=C(X_0 ,0;K_1 ,T_1),$$
which is a standard European call. This is the case illustrated by
Figure \ref{Call_fig}e.

\item If $I_1<K_1$, then the nonlinear equation
$f_2^C(x)=0$ (for $x>K_1$) has either one solution denoted as $I_2$
or none as illustrated by Figures  \ref{Call_fig}a and
\ref{Call_fig}b respectively. In the case of one solution $I_2$, the
call option is extended when $I_1<X_{T_1}<I_2$; exercised when
$X_{T_1}\ge I_2$; and expires worthless when $X_{T_1}\le I_1$. If
there is no solution, then the call option is extended when
$I_1<X_{T_1}$ and expires worthless when $I_1\ge X_{T_1}$. In
particular:
\begin{itemize}
\item If $q_{12}>0$, then there is finite $I_2$.
\item If $q_{12}=0$, then $I_2$ is finite when $K_1-A-K_2e^{-r_{12}(T_2-T_1)}<
0$;
and $f_2^C(x)=0$ has no finite solution when
$K_1-A-K_2e^{-r_{12}(T_2-T_1)}\ge 0$.
\end{itemize}

Then, the today's price of the holder-extendible call can be
calculated by integrating (\ref{call_intergal_solution_general_eq})
with $\Omega_C=[I_1,I_2]$ and $\overline\Omega_C=[I_2,\infty)$ (the
case when $f_2^C(x)=0$ has no solution can be treated by setting
$I_2=\infty$) to obtain
\begin{eqnarray}
\label{callformula_2solutions_eq}
 Q_C (X_0 ,0;K_1,K_2,T_1,T_2) &=& C(X_0 ,0;K_1 ,T_1 )\nonumber\\
 &&+ X_0 e^{(\mu _{02}-r_{02}) T_2 }M_2\left(-\widetilde{g}_1(I_2),-\widetilde{g}_1(I_1), - \infty ,
-\widetilde{g}_2(K_2);\rho\right) \nonumber\\
 &&- K_2e^{ - r_{02} T_2 } M_2(-g_1(I_2),-g_1(I_1), - \infty ,-g_2(K_2);\rho ) \nonumber\\
 && - Ae^{ - r_{01}T_1 } M(
-g_1(I_2),-g_1(I_1))\nonumber\\
 && - X_0 e^{-q_{01} T_1 }M(-\widetilde{g}_1(I_2),
 -\widetilde{g}_1(K_1))
\nonumber\\
&&+ K_1e^{ - r_{01} T_1 } M(-g_1(I_2),-g_1(K_1)).
 \end{eqnarray}
\noindent This expression will reduce to the original Longstaff
(1990, equation 7) formula for the holder-extendible call after
setting (\ref{comparison_setting_eq}) and Longstaff (1990) notation
\begin{equation}\label{gamma_const_call_eq} \gamma _1 = -\widetilde{g}_1(I_2),
\quad\gamma _2 = -\widetilde{g}_1(I_1), \quad\gamma _3 =
-\widetilde{g}_2(K_2), \quad\gamma _4 = -\widetilde{g}_1(K_1).
\end{equation}

\end{itemize}

\example{Consider the holder-extendible call with initial maturity
$T_1=1$ year that can be extended to $T_2=2$ years. The model
parameters are: spot $X_0=0.9$, strike on decision $K_1=0.9$, strike
at final maturity $K_2=0.95$, interest rate $r=0.02$, ``dividend"
$q=0$, volatility $\sigma=0.3$, extra premium $A=0.03$. The payoffs
and critical values are shown in Figure \ref{Call_fig}a. Solving
nonlinear equations (\ref{call_crit_equation1_eq}) and
(\ref{call_crit_equation2_eq}) via bisection algorithm gives
$I_1\approx 0.734$ and $I_2\approx 1.074$. Finally using formula
(\ref{callformula_2solutions_eq}), find that the today's price of
the holder-extendible call is $Q_C (X_0 ,0;K_1,K_2,T_1,T_2)\approx
0.129$.}

\subsection{Negative ``dividend''}\label{Call_negdiv_sec}
The case of negative ``dividend", $q_{12}<0$, is a bit complicated.
This is because the first derivative of European call $\Delta_C(x)$,
see (\ref{CallDelta_eq}), can be become larger than 1.

\begin{itemize}
\item If $I_1> K_1$, then nonlinear equation (\ref{call_crit_equation2_eq})
has one finite solution $I_2$, and the call is extended when
$X_{T_1}>I_2$; this case is shown in Figure \ref{Call_fig}d. The
price is calculated by integrating
(\ref{call_intergal_solution_general_eq}) with
$\Omega_C=[I_2,\infty)$ and $\overline\Omega=[K_1,I_2]$
\begin{eqnarray}
\label{callformula_1solution_eq}
 Q_C (X_0 ,0;K_1,K_2,T_1,T_2) &=& C(X_0 ,0;K_1 ,T_1 )+ X_0 e^{-q_{02} T_2 }
 N_2(-\widetilde{g}_1(I_2),-\widetilde{g}_2(K_2);\rho ) \nonumber\\
 &&- K_2e^{ - r_{02} T_2 } N_2(-g_1(I_2),-g_2(K_2);\rho)  -
 Ae^{ - r_{01}T_1 } N(-g_1(I_2) ) \nonumber\\
 && - X_0 e^{-q_{01} T_1 }N(-\widetilde{g}_1(I_2)) +
 K_1e^{ - r_{01} T_1 } N(-g_1(I_2) ).
 \end{eqnarray}
 \item If $I_1 \le K_1$, then nonlinear equation
(\ref{call_crit_equation2_eq}) has either two solutions $I_2$ and
$I_3$ (with $I_3\ge I_2\ge I_1$, as illustrated on Figure
\ref{Call_fig}c, and the call is extended if $I_1<X_{T_1}<I_2$ or
$X_{T_1}>I_3$) or none as illustrated on Figure \ref{Call_fig}b. For
the latter, the call is extended if $X_{T_1}>I_1$. Specifically,
$f_2^C(x)$ has a minimum at $x=x^\ast$ where $df_2^C(x)/dx=0$. Using
(\ref{CallDelta_eq}), it is easy to find that
$$
f_2^C(x^\ast)=K_1-A-K_2e^{-r_{12}(T_2-T_1)}N(F_{N}^{-1}(e^{q_{12}(T_2-T_1)})).
$$
Thus, if $f_2^C(x^\ast)<0$ then there are two finite solutions,
otherwise there is no solution. In the case of no solution the price
is given by (\ref{callformula_2solutions_eq}) with $I_2$ set to
$\infty$. In the case of two solutions, integration
(\ref{call_intergal_solution_general_eq}) with $\Omega_C=
\{(I_1,I_2) \bigcup (I_3,\infty)\}$ and $\overline\Omega_C=
[I_2;I_3]$ gives
\begin{eqnarray} \label{callformula_3solution_eq}
&& \hspace{-1cm}Q_C (X_0 ,0;K_1,K_2,T_1,T_2) = C(X_0 ,0;K_1 ,T_1 )\nonumber\\
&& + X_0 e^{-q_{02} T_2 }
\left[M_2(-\widetilde{g}_1(I_2),-\widetilde{g}_1(I_1),-\infty,-\widetilde{g}_2(K_2);\rho
)+
N_2(-\widetilde{g}_1(I_3),-\widetilde{g}_2(K_2);\rho )\right] \nonumber\\
 && - K_2e^{ - r_{02} T_2 } \left[M_2(-g_1(I_2),-g_1(I_1),-\infty,-g_2(K_2);\rho)+N_2(-g_1(I_3),-g_2(K_2);\rho)\right] \nonumber\\
 && - Ae^{ - r_{01}T_1 } \left[M(g_1(I_1),g_1(I_2)) + N(-g_1(I_3))\right]\nonumber\\
 && + X_0 e^{-q_{01} T_1 }\left[M(\widetilde{g}_1(I_2),\widetilde{g}_1(I_3))-
 N(-\widetilde{g}_1(K_1))\right]\nonumber\\
 && -  K_1e^{ - r_{01} T_1 } \left[M(g_1(I_2),g_1(I_3))-N(-g_1(K_1))\right].
 \end{eqnarray}

 \end{itemize}

\example{Consider the holder-extendible call with initial maturity
$T_1=1$ year that can be extended to $T_2=2$ years. The model
parameters are: spot $X_0=0.9$, strike on decision $K_1=0.9$, strike
at final maturity $K_2=1.4$, interest rate $r=0.02$, ``dividend"
$q=-0.28$, volatility $\sigma=0.3$, extra premium $A=0.03$. The
payoffs and critical values are shown in Figure \ref{Call_fig}c.
Solving nonlinear equations (\ref{call_crit_equation1_eq}) and
(\ref{call_crit_equation2_eq}) via bisection algorithm gives
$I_1\approx 0.771$, $I_2\approx 1.024$ and $I_3\approx 1.459$.
Finally using formula (\ref{callformula_3solution_eq}), find that
the today's price of the holder-extendible call is $Q_C (X_0
,0;K_1,K_2,T_1,T_2)\approx 0.357$.}

\section{Holder-extendible put}
\begin{center}
\begin{figure}[p]
\centerline{\includegraphics[scale=0.65]{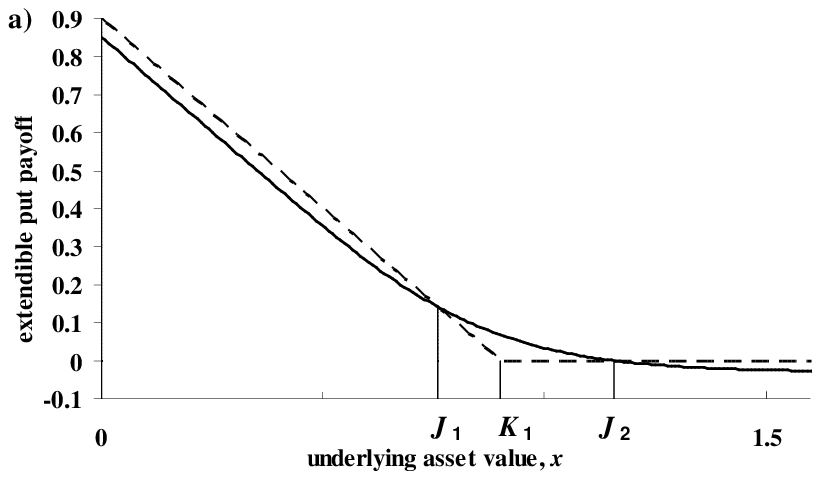}\hspace{0.2cm}\includegraphics[scale=0.65]{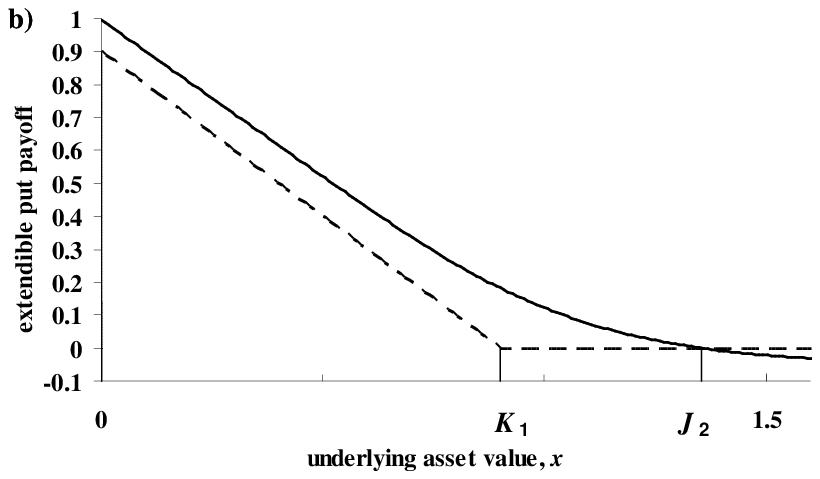}}
\vspace{0.5cm}
\centerline{\includegraphics[scale=0.65]{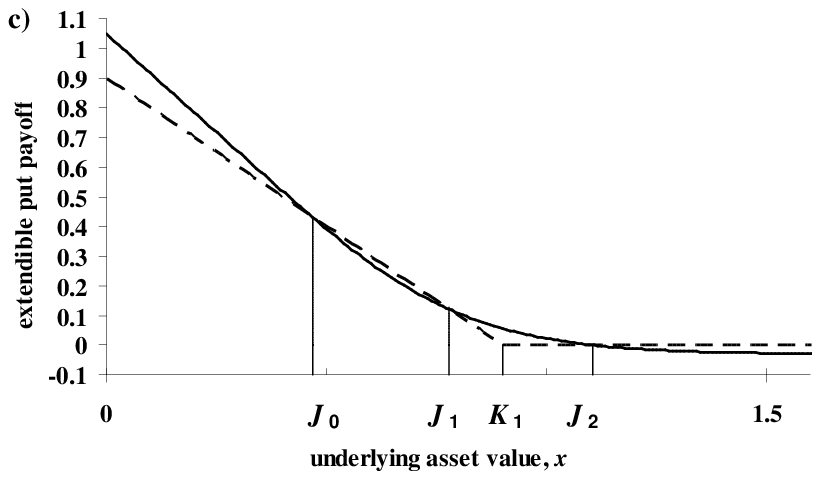}\hspace{0.2cm}\includegraphics[scale=0.65]{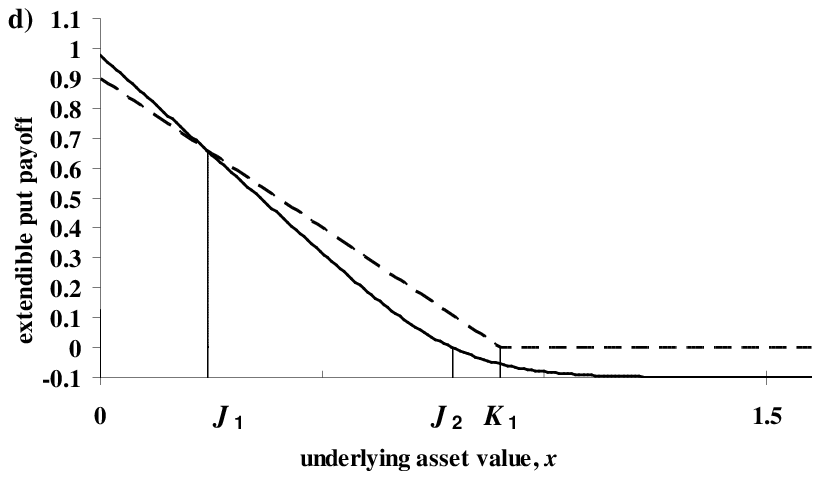}}
\vspace{0.5cm}
\centerline{\includegraphics[scale=0.65]{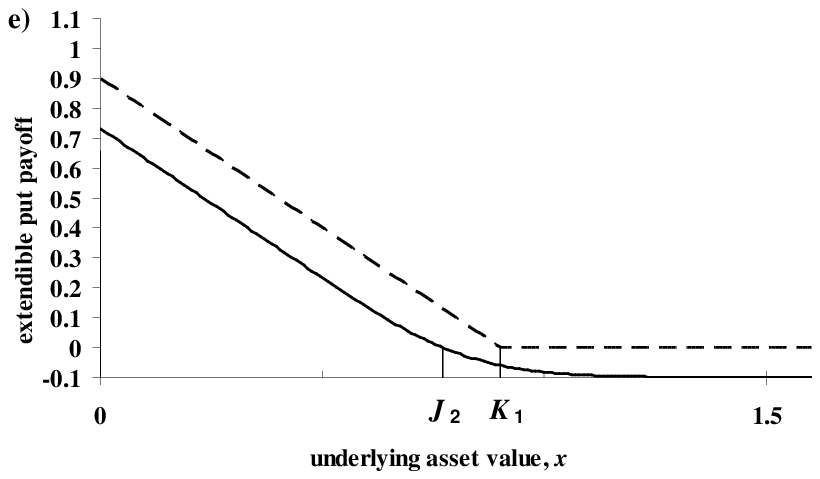}}
\caption{Possible holder-extendible put payoffs on a decision time
$T_1$. The payoff is determined by choosing the largest value
between solid line $P(x,T_1;K_2,T_2) - A$ and dashed line $\max
\left(K_1-x,0\right)$.} \label{Put_fig}
\end{figure}
\end{center}

The decision at $T_1 $ to extend or exercise the put option is
determined by comparing two risky payoffs
\begin{equation}
P(X_{T_1 } ,T_1;K_2,T_2) - A\quad\mbox{and}\quad \max
\left(K_1-X_{T_1 }
 ,0\right);
\end{equation}
\noindent and choosing the largest payoff. If the first payoff is
larger then the option is extended, otherwise it is exercised when
$X_{T_1 } < K_1$ or expires worthless when $X_{T_1 } \ge K_1$; for
an illustrative example, see Figure \ref{Put_fig}. Note that the
standard European put $P(x ,T_1; K_2 ,T_2)$ is calculated at time
$T_1$ for maturity at $T_2$.

Denote the region of $X_{T_1}$ values where the put option is
extended as

\begin{equation}\Omega_P=\{x\ge 0: P(x ,T_1;K_2,T_2) - A > \max \left(K_1-x
,0\right)\} \end{equation}

\noindent and the region where it is exercised as

\begin{equation}\overline{\Omega}_P=\{0\le x< K_1: K_1-x\ge P(x
,T_1;K_2,T_2) - A \}. \end{equation}

\noindent For all other values of $X_{T_1}$, the option expires
worthless. Then the holder-extendible put price can be written as
\begin{eqnarray}
\label{put_intergal_solution_general_eq}
 &&Q_P (X_0 ,0;K_1,K_2,T_1,T_2) = e^{ - r_{01} T_1 }\int_{ - \infty }^\infty  \max
\left(P(x_1,T_1;K_2 ,T_2) - A,K_1-x_1
,0\right){n({z}_1)d{z}_1} \nonumber\\
&&\hspace{2.0cm}=e^{ - r_{02} T_2 }\int_{{x}_1\in\Omega_P}
 {d{z}_1\int_{-\infty}^{g_2(K_2)} {(K_2-x_2)n_2({z}_1,{z}_2;\rho )d{z}_2} } \nonumber\\
 &&\hspace{2.5cm} - e^{ - r_{01} T_1 }A\int_{{x}_1\in\Omega_P}n({z}_1)d{z}_1
+  e^{ - r_{01} T_1 }\int_{ x_1\in\overline\Omega_P}
n({z}_1)(K_1-x_1)d{z}_1.
 \end{eqnarray}

\noindent Here $x_1=g^{-1}_1({z}_1)$ and $x_2=g^{-1}_2({z}_2)$ are
functions of ${z}_1$ and ${z}_2$ as given by
(\ref{invtransform_func}).

 The regions
$\Omega_P$ and $\overline{\Omega}_P$ can be determined using
solutions (critical asset values) of nonlinear equations
\begin{equation}
\label{put_crit_equation1_eq} f_1^P(x)= P(x ,T_1;K_2 ,T_2) - K_1+x-
A=0,\quad 0\le x< K_1,
\end{equation}
\noindent and
\begin{equation}
\label{put_crit_equation2_eq} f_2^P(x)=P(x ,T_1; K_2 ,T_2) - A=0,\quad x\ge 0.\\
\end{equation}

\noindent As in the case of the holder-extendible call, these can be
solved numerically using the Newton-Raphson algorithm combined with
the standard bisection algorithm to avoid numerical difficulties
when corresponding derivatives are close to zero.

If $A>P(0,T_1;K_2,T_2)=K_2e^{-r_{12}(T_2-T_1)}$, then $f_2^P(x)<0$
for all $x\ge 0$ and thus put is never extended, i.e.
$Q_P(X_0,0;K_1,K_2,T_1,T_2)=P(X_0 ,0;K_1 ,T_1)$. Otherwise,
$f_2^P(x)=0$ has one solution, denoted as $x=J_2$, and this case is
considered hereafter.

The first equation $f_1^P(x)=0$ may have no solutions, one solution
(denoted as $J_1$) or two solutions (denoted as $J_0$ and $J_1$)
depending on the option characteristics (strikes, maturities, model
parameters), that will determine the today's option price; Figure
\ref{Put_fig} illustrates possible cases. Below we consider the
cases of non-negative and negative ``dividend", i.e. the cases
$q_{12}\ge 0$ and $q_{12}<0$ respectively. Similar to the
holder-extendible call, $f_1^P(x)=0$ may have one or no solution if
$q_{12}\ge 0$ and two solutions if $q_{12}<0$; Longstaff (1990)
considers the case of zero ``dividend" $q_{12}=0$.

All conditions, listed in Sections \ref{Put_nonnegdiv_sec} and
\ref{Put_negdiv_sec}, on option characteristics to determine
solution type can easily be proved using the facts that the European
put price $P(x,T_1;K_2,T_2)$ is continuous and decreasing function
of $x$, and its first derivative
\begin{equation}
\Delta_P(x)=\frac{\partial P(x,T_1;K_2,T_2)}{\partial
x}=e^{-q_{12}(T_2-T_1)}\left(N\left(\frac{\ln(x/K)+(\mu_{12}+\frac{1}{2}\sigma^2_{12})(T_2-T_1)}{\sigma_{12}\sqrt{T_2-T_1}}\right)-1\right)
\label{PutDelta_eq}
\end{equation}
is negative. It is important to note that  $-1\le \Delta_P(x)\le 0$
when $q_{12}\ge 0$; however, if $q_{12}< 0$ then $\Delta_P(x)<-1$ is
possible.

\subsection{Non-negative ``dividend"}\label{Put_nonnegdiv_sec}
Here we consider the case of non-negative ``dividend" $q_{12}\ge 0$.
\begin{itemize}

\item If $J_2\le K_1$ then the put is never extended and thus
$$Q_P(X_0 ,0;K_1,K_2,T_1,T_2)=P(X_0 ,0;K_1 ,T_1),$$
which is a standard European put; this case is shown in Figure
\ref{Put_fig}e.

\item If $J_2>K_1$, nonlinear equation
(\ref{put_crit_equation1_eq}) may have one solution $J_1$ (i.e. the
option is extended if $J_1<X_{T_1}<J_2$) or none as shown in Figure
\ref{Put_fig}a and Figure \ref{Put_fig}b respectively. The latter
case corresponds to put option which is extended for $X_{T_1}<J_2$.
In particular,  if $K_1<K_2e^{-r_{12}(T_2-T_1)}-A$ then there are no
solutions, otherwise there is one finite solution $J_1$. Then the
today's price can be calculated by integrating
(\ref{put_intergal_solution_general_eq}) with $\Omega_P=[J_1,J_2]$
and $\overline{\Omega}_P=[0,J_1]$ (the case when $f_1^P(x)$ has no
solution can be treated by setting $J_1=0$) to obtain
\begin{eqnarray}
\label{putformula_2solutions_eq}
Q_P (X_0 ,0;K_1,K_2,T_1,T_2)&=&P(X_0,0;K_1 ,T_1 )\nonumber\\
&& - X_0 e^{(\mu _{02}-r_{02}) T_2 }M_2(
\widetilde{g}_1(J_1),\widetilde{g}_1(J_2),-\infty,
\widetilde{g}_2(K_2);\rho )  \nonumber\\
 &&+ K_2e^{ - r_{02} T_2 } M_2(g_1(J_1)
,g_1(J_2),-\infty,
g_2(K_2) ;\rho ) \nonumber \\
 &&- Ae^{ - r_{01} T_1 } M(-g_1(J_2),-g_1(J_1))  \nonumber\\
  &&+ X_0 e^{(\mu _{01}-r_{01}) T_1 }M(-\widetilde{g}_1(K_1),-\widetilde{g}_1(J_1)) \nonumber \\
 &&- K_1e^{ - r_{01} T_1 } M( -g_1(K_1),-g_1(J_1) ).
 \end{eqnarray}
\noindent

 This formula appeared in the literature with erroneous typographical errors. To make a comparison easier,
 re-write the formula using Longstaff (1990) notation
\begin{equation}\label{gamma_const_put_eq} \gamma _1 = -\widetilde{g}_1(J_2),
\quad\gamma _2 = - \widetilde{g}_1(J_1), \quad\gamma _3 =
\widetilde{g}_2(K_2) ,\quad\gamma _4 = -\widetilde{g}_1(K_1).
\end{equation}

Then the holder-extendible put can be written as
\begin{eqnarray}
&& Q_P (X_0 ,0;K_1,K_2,T_1,T_2) =P(X_0,0;K_1 ,T_1 )\nonumber\\
 &&\hspace{2cm}- \underline{X_0 e^{(\mu _{02}-r_{02}) T_2 }M_2( - \gamma _2 , - \gamma
_1 , - \infty , - \gamma _3;\rho )} \nonumber \\
 && \hspace{2cm}+\underline{K_2e^{ - r_{02} T_2 } M_2(\sigma _{01} \sqrt {T_1 } - \gamma _2
,\sigma _{02} \sqrt {T_2 } - \gamma _1 , - \infty ,\sigma _{02}
\sqrt {T_2 } -
\gamma _3 ;\rho )}\nonumber\\
 &&\hspace{2cm} - Ae^{ - r_{01} T_1 }M(\gamma _1 - \sigma _{01} \sqrt {T_1 } ,\gamma
_2 - \sigma _{01} \sqrt {T_1 } ) \nonumber\\
 &&\hspace{2cm} + X_0 e^{(\mu _{01}-r_{01}) T_1 }M(\gamma _4 ,\gamma _2 )\nonumber \\
 &&\hspace{2cm} - K_1e^{ - r_{01} T_1 } M(\gamma _4 - \sigma _{01} \sqrt {T_1 }
,\gamma _2 - \sigma _{01} \sqrt {T_1 } ).
 \end{eqnarray}
After setting (\ref{comparison_setting_eq}), the difference between
this formula (see underlined terms) and Longstaff (1990, equation
12) is clear. For the latter: $\gamma_3$, $\gamma _3 - \sigma \sqrt
{T_2 } $ and $\rho $ should be changed to $-\gamma _3$, $ - \gamma
_3 + \sigma \sqrt {T_2 } $ and $ - \rho $ respectively; also the
factor in the 3rd term, $\exp ( - r(T_2 - T_1 ))$, should be
replaced with $\exp ( - rT_2 )$. Also, note that the formula for the
holder-extendible put in Haug (1998, equation 2.15, p.48) also has a
typographical error where $\rho $ should be changed to $ - \rho $.
When comparing the formulas the following symmetry property is
useful: $M_2(a,b,c,d,\rho ) = M_2( - b, - a,c,d, - \rho )$.

\end{itemize}

\example{Consider the holder-extendible put with initial maturity
$T_1=1$ year that can be extended to $T_2=2$ years. The model
parameters are: spot $X_0=0.9$, strike on decision $K_1=0.9$, strike
at final maturity $K_2=0.9$, interest rate $r=0.02$, ``dividend"
$q=0$, volatility $\sigma=0.3$, extra premium $A=0.03$. The payoffs
and critical values are shown in Figure \ref{Put_fig}a. Solving
nonlinear equations (\ref{put_crit_equation1_eq}) and
(\ref{put_crit_equation2_eq}) via bisection algorithm gives
$J_1\approx 0.758$ and $J_2\approx 1.157$. Finally using formula
(\ref{putformula_2solutions_eq}), find that the today's price of the
holder-extendible put is $Q_P (X_0 ,0;K_1,K_2,T_1,T_2)\approx
0.113$.}

\subsection{Negative ``dividend''}\label{Put_negdiv_sec}
The case of negative ``dividend", $q_{12}<0$, is a bit complicated
due to the fact that $\Delta_P$ may become less than -1; see
(\ref{PutDelta_eq}).

\begin{itemize}
\item If $J_2<K_1$ then nonlinear equation (\ref{put_crit_equation1_eq})
has either one finite solution $J_1$, and the call is extended if
$X_{T_1}<J_1$, or none; these cases are shown in Figure
\ref{Put_fig}d and Figure \ref{Put_fig}e respectively. Specifically,
if $K_1<K_2e^{-r_{12}(T_2-T_1)}-A$ then there is one solution,
otherwise there is no solution. If there is no solution then the put
is never extended $Q_P(X_0 ,0;K_1,K_2,T_1,T_2)=P(X_0 ,0;K_1 ,T_1)$.
Otherwise, the price is calculated by integrating
(\ref{put_intergal_solution_general_eq}) with $\Omega_P=[0,J_1]$ and
$\overline\Omega_P=[J_1,K_1]$
\begin{eqnarray}
\label{putformula_1solution_eq}
 Q_P (X_0 ,0;K_1,K_2,T_1,T_2) &=& P(X_0 ,0;K_1 ,T_1 )- X_0 e^{-q_{02} T_2 }
 N_2(\widetilde{g}_1(J_1) ,\widetilde{g}_2(K_2);\rho ) \nonumber\\
 &&+ K_2e^{ - r_{02} T_2 } N_2(g_1(J_1),g_2(K_2);\rho)  -
 Ae^{ - r_{01}T_1 } N(g_1(J_1) ) \nonumber\\
 && + X_0 e^{-q_{01} T_1 }N(\widetilde{g}_1(J_1) ) -
 K_1e^{ - r_{01} T_1 } N(g_1(J_1) ).
 \end{eqnarray}
 \item If $J_2\ge  K_1$, then nonlinear equation
(\ref{put_crit_equation1_eq}) has either two solutions $J_0$ and
$J_1$ (put is extended if $0<X_{T_1}<J_0$ or $J_1<X_{T_1}<J_2$); one
solution (put is extended if $J_1<X_{T_1}<J_2$) or none (put is
extended if $X_{T_1}>J_1$); these three cases are shown in Figure
\ref{Put_fig}c, \ref{Put_fig}a and \ref{Put_fig}b respectively.
Specifically, $f_1^P(x)$ has a minimum at $x=x^\ast$ where
$df_1^P(x)/dx=0$. Using (\ref{PutDelta_eq}), it is easy to find that
$$
f_1^P(x^\ast)=K_2e^{-r_{12}(T_2-T_1)}N(\sigma_{12}\sqrt{T_2-T_1}-d)-A-K_1,\quad
d=F_N^{-1}\left(1-e^{q_{12}(T_2-T_1)}\right).
$$
Thus, if $f_1^P(x^\ast)>0$ then there is no solution and price can
be calculated via (\ref{putformula_2solutions_eq}) with $J_1$ set to
zero; if $f_1^P(x^\ast)\le 0$ and $K_1>K_2e^{-r_{12}(T_2-T_1)}$ then
there is one finite solution $J_1$ and price can be calculated using
(\ref{putformula_2solutions_eq}); if $f_1^P(x^\ast)\le 0$ and
$K_1\le K_2e^{-r_{12}(T_2-T_1)}$ then there are two finite solutions
$J_0\le J_1$. For the last case, integration
(\ref{put_intergal_solution_general_eq}) with $\Omega= \{[0,J_0)
\bigcup (J_1,J_2)\}$ and $\overline\Omega_P= [J_0,J_1]$ gives
\begin{eqnarray}
\label{putformula_3solution_eq}
&& Q_P (X_0 ,0;K_1,K_2,T_1,T_2) = P(X_0 ,0;K_1 ,T_1 )\nonumber\\
&&\hspace{2cm} - X_0 e^{-q_{02} T_2 } \left[N_2(\widetilde{g}_1(J_0)
,\widetilde{g}_2(K_2);\rho ) +
M_2(\widetilde{g}_1(J_1),\widetilde{g}_1(J_2)
,-\infty,\widetilde{g}_2(K_2);\rho )\right] \nonumber\\
 &&\hspace{2cm}+
 K_2e^{ - r_{02} T_2 } \left[N_2(g_1(J_0),g_2(K_2);\rho)+
 M_2(g_1(J_1),g_1(J_2),-\infty,g_2(K_2);\rho)\right] \nonumber\\
 &&\hspace{2cm} - Ae^{ - r_{01}T_1 } \left[N(g_1(J_0)) +M(g_1(J_1),g_1(J_2))\right]\nonumber\\
 &&\hspace{2cm} - X_0 e^{-q_{01} T_1 }\left[M(\widetilde{g}_1(J_0),\widetilde{g}_1(J_1))-
 N(\widetilde{g}_1(K_1) )\right]\nonumber\\
 &&\hspace{2cm} +  K_1e^{ - r_{01} T_1 } \left[M(g_1(J_0),g_1(J_1))-N(g_1(K_1))\right].
 \end{eqnarray}
 \end{itemize}

\example{Consider the holder-extendible put with initial maturity
$T_1=1$ year that can be extended to $T_2=2$ years. The model
parameters are: spot $X_0=0.9$, strike on decision $K_1=0.9$, strike
at final maturity $K_2=1.1$, interest rate $r=0.02$, ``dividend"
$q=-0.28$, volatility $\sigma=0.3$, extra premium $A=0.03$. The
payoffs and critical values are shown in Figure \ref{Call_fig}c.
Solving nonlinear equations (\ref{put_crit_equation1_eq}) and
(\ref{put_crit_equation2_eq}) via bisection algorithm gives
$J_0\approx 0.468$, $J_1\approx 0.779$ and $J_2\approx 1.107$.
Finally using formula (\ref{putformula_3solution_eq}), find that the
today's price of the holder-extendible put is $Q_P (X_0
,0;K_1,K_2,T_1,T_2)\approx 0.034$.}

\section{Conclusion}
We have derived closed-form formulas for the holder-extendible call
and put in a presence of ``dividend" yield that can be not only zero
or positive but may also be negative. The last corresponds to the
case of negative foreign interest rate for FX options or the assets
with storage costs; it may also appear when transaction costs are
accounted for or for real options where drift is larger than
interest rate. The considered case is more general than zero
``dividend" case studied in Longstaff (1990) and Chung and Johnson
(2011) or non-negative ``dividend" case treated in Haug (2000) and
Chateau and Wu (2007). It is important to note that negative
``dividend" may lead to solutions involving three critical asset
levels defining decision regions while non-negative ``dividend" case
leads to solutions involving only two critical levels. Also, we
fixed the erroneous typos for the holder-extendible put formula
published in previous literature. Finally, all formulas are derived
for the case of geometric Brownian motion with the time-dependent
drift and volatility which is important to use the solution in
practical applications.

\section{Appendix}
All integrals involved into calculation of the today's option price
(\ref{call_intergal_solution_general_eq}) and
(\ref{put_intergal_solution_general_eq}) can be found using
closed-form integrals
\begin{eqnarray}
\label{useful_identity_eq}
\begin{array}{l}
\int_{ - \infty }^a {\int_{ - \infty }^b {n_2(x,y;\rho )} } e^{\beta
x}dxdy = \exp \left( {\frac{\beta ^2}{2}} \right)N_2(a
- \beta ,b - \beta \rho ;\rho ); \\
 \int_a^\infty {\int_b^\infty {n_2(x,y;\rho )} } e^{\beta x}dxdy =
\exp \left( {\frac{\beta ^2}{2}} \right)N_2(\beta - a,\beta \rho -
b;\rho ); \\
 \int_a^\infty {n(x)} e^{\beta x}dx = \exp \left( {\frac{\beta
^2}{2}} \right)N(\beta - a); \\
\int_{ - \infty }^a {n(x)} e^{\beta x}dx = \exp \left( {\frac{\beta
^2}{2}} \right)N(a - \beta ).\\
\end{array}
 \end{eqnarray}

\noindent Also, the following relationships for the probability
functions are used throughout the paper to simplify the formulas

\begin{equation}
\begin{array}{l}
\label{normprob_def_eq}
 M_2(a,b,c,d;\rho ) = N_2(b,d;\rho ) - N_2(a,d;\rho ) - N_2(b,c;\rho ) + N_2(a,c;\rho
);\\
M_2(a,b,-\infty,d;\rho ) = N_2(b,d;\rho ) - N_2(a,d;\rho );\\
M(a,b) = N(b) - N(a).\\
\end{array}
\end{equation}


\begin{thebibliography}{10}

\bibitem{Longstaff1990} Longstaff, F.A. (1990). Pricing Options with Extendible Maturities: Analysis
and Applications, \textit{Journal of Finance}, \textbf{45}(3),
935-957.


\bibitem{Haug1998} Haug, E.G. (1998). \textit{The Complete Guide to Options Pricing Formulas},
McGraw-Hill.

\bibitem{JPChateau2007} Chateau, J.-P. and Wu, J. (2007). Basel-2
capital adequacy: computing the `fair' capital charge for loan
commitment `true' credit risk. \textit{International Review of
Financial Analysis}, \textbf{16}, 1-21.

\bibitem{ChJo2010} Chung, Y.P. and Johnson, H. (2011). Extendible
options: The general case, \textit{Finance Research Letters},
\textbf{8}(1), 15-20.
\end{thebibliography}
\end{document}